\documentclass{PoS}
\usepackage{wrapfig}

\title{Impact of LHC top-quark pair measurements to CTEQ-TEA PDF analysis}

\ShortTitle{Impact of $t\bar{t}$ pair measurements to CT PDFs}

\author{Orkash Amat\\
    School of Physics Science and Technology, Xinjiang University, Urumqi, Xinjiang 830046 China\\
    E-mail:\email{orkaxmamat@163.com}}

\author{Micha\l{} Czakon\\
    Institut f\"ur Theoretische Teilchenphysik und Kosmologie, RWTH Aachen University, D-52056 Aachen, Germany\\
    E-mail:\email{mczakon@physik.rwth-aachen.de}}

\author{Sayipjamal Dulat\\
    School of Physics Science and Technology \& Center for Theoretical Physics, \\ 
    Xinjiang University, Urumqi, Xinjiang 830046 China\\
    E-mail: \email{sdulat@hotmail.com}}

\author{\speaker{Tie-Jiun Hou}\\
    Department of Physics, College of Sciences, Northeastern University, Shenyang 110819, China\\
    E-mail: \email{houtiejiun@mail.neu.edu.cn}}

\author{Joey Huston\\
    Department of Physics and Astronomy, Michigan State University, East Lansing, MI 48824 U.S.A.\\
    E-mail: \email{Huston@pa.msu.edu}}

\author{Alexander Mitov\\
    Cavendish Laboratory, University of Cambridge, Cambridge CB3 0HE, UK\\
    E-mail:\email{adm74@cam.ac.uk}}

\author{Andrew S.~Papanastasiou\\
	Cavendish Laboratory, University of Cambridge, Cambridge CB3 0HE, UK\\
	MRC Institute of Genetics and Molecular Medicine, University of Edinburgh, Crewe Road, Edinburgh EH4 2XU, UK\\
	E-mail:\email{apapanas@inf.ed.ac.uk}}

\author{Carl Schmidt\\
    Department of Physics and Astronomy, Michigan State University, East Lansing, MI 48824 U.S.A.\\
    E-mail: \email{schmidt@pa.msu.edu}}

\author{Ibrahim Sitiwaldi\\
    School of Physics Science and Technology \& Center for Theoretical Physics, \\ 
    Xinjiang University, Urumqi, Xinjiang 830046 China\\
    E-mail: \email{ibrahim010@sina.com}}

\author{Keping Xie\\
    Department of Physics, Southern Methodist University, Dallas, TX 75275-0181, U.S.A.\\
    E-mail: \email{kepingx@mail.smu.edu}}

\author{Zhite Yu\\
    Department of Physics and Astronomy, Michigan State University, East Lansing, MI 48824 U.S.A.\\
    E-mail: \email{yuzhite@msu.edu}}

\author{C.-P. Yuan\\
    Department of Physics and Astronomy, Michigan State University, East Lansing, MI 48824 U.S.A.\\
    E-mail: \email{yuan@pa.msu.edu}}

\abstract{Detailed studies have been carried out on the impact of the LHC top quark pair production data 
          on gluon PDF, in the context of the CTEQ-TEA global PDF fit, with the ePump-updating method. 
          The considered $t\bar{t}$ data include single differential distributions from ATLAS and double 
          differential distributions from CMS, both at 8 TeV.  All analyses 
          have been carried out at the NNLO, using fastNNLO tables.
 We show that the sensitivity per data point of the LHC $t\bar{t}$ data is 
 similar to that of jet data, as   included in the CT14HERA2 fit, while the total sensitivity of the  
 present $t\bar{t}$ data is not as large as the jet data because of the much smaller number of $t\bar{t}$ data points in the presently available data. 
}

\dedicated{Cavendish-HEP-19/14, MSUHEP-19-018}

\FullConference{XXVII International Workshop on Deep-Inelastic Scattering and Related Subjects - DIS2019\\
        8-12 April, 2019\\
        Torino, Italy}

\begin{document}

The top-quark pair production is a brand new observable available 
for global analysis in CTEQ-TEA PDFs after CT14HERA2~\cite{Hou:2016nqm}.
For experimental side, we consider the absolute and normalized 
one-dimensional $p_T$, $y_t$, $m_{t\bar{t}}$ and $y_{t\bar{t}}$ distributions  
from ALTAS~\cite{Aad:2015mbv} and CMS~\cite{Khachatryan:2015oqa}, 
and the two-dimensional distributions from CMS~\cite{Sirunyan:2017azo}.
Theory prediction is done at the NNLO QCD with $\mu_R,\mu_f = \frac{H_T}{4}$ or $\frac{m_T}{4}$ 
 through fastNLO 
grids~\cite{Czakon:2016dgf, Czakon:2017dip}.
Instead of implementing in real global analysis, we study the 
impact of top-quark pair production on PDFs in the framework of 
CT14HERA2 by using ePump (Error PDF Updating Method Package)~\cite{Schmidt:2018hvu}.

%
\begin{table}[htbp]
\begin{center}
\begin{tabular}{| c | c c | c | c |}
\hline
 Observable & Detector & & Npts & $\chi^2/N$ \\
\hline
 inclusive jet                                    & CDF        &  ~\cite{Aaltonen:2008eq}              & 72   & 1.50 \\            
\hline                                                                                                     
 inclusive jet                                    & D0         &  ~\cite{Abazov:2008ae}                & 110  & 1.03 \\            
\hline                                                                                                     
 inclusive jet                                    & ATLAS      &  ~\cite{Aad:2011fc}                   & 90   & 0.57 \\            
\hline                                                                                                     
 inclusive jet                                    & CMS        & ~\cite{Chatrchyan:2012bja}            & 133  & 0.93 \\            
\hline                                                               
 $\frac{1}{\sigma}  \frac{d\sigma}{dp^t_T}$       & ATLAS, CMS &~\cite{Aad:2015mbv,Khachatryan:2015oqa}& 8,8  & 0.39, 3.88 \\
\hline                                                               
 $\frac{1}{\sigma}  \frac{d\sigma}{dy_t}$         & ATLAS, CMS &~\cite{Aad:2015mbv,Khachatryan:2015oqa}& 5,10 & 2.70, 2.53\\
\hline                                                               
 $\frac{1}{\sigma}  \frac{d\sigma}{dm_{t\bar t}}$ & ATLAS, CMS &~\cite{Aad:2015mbv,Khachatryan:2015oqa}& 7,7  & 0.25, 8.67\\
\hline                                                               
 $\frac{1}{\sigma}  \frac{d\sigma}{dy_{t\bar t}}$ & ATLAS, CMS &~\cite{Aad:2015mbv,Khachatryan:2015oqa}& 5,10 & 2.46, 3.67\\
\hline                                                               
 $  \frac{d\sigma}{dp^t_T}$                       & ATLAS      &~\cite{Aad:2015mbv}                    & 8    & 0.34 \\
\hline                                                                                                 
 $  \frac{d\sigma}{dy_t}$                         & ATLAS      &~\cite{Aad:2015mbv}                    & 5    & 3.18 \\
\hline                                                                                                 
 $  \frac{d\sigma}{dm_{t\bar t}}$                 & ATLAS      &~\cite{Aad:2015mbv}                    & 7    & 0.45\\
\hline                                                                                                 
 $  \frac{d\sigma}{dy_{t\bar t}}$                 & ATLAS      &~\cite{Aad:2015mbv}                    & 5    & 4.65\\
\hline                                                               
 $d^2\sigma/dy_t dp^t_T$                          & CMS        &~\cite{Sirunyan:2017azo}               & 16   & 1.23 \\
\hline                                                                                                 
 $d^2\sigma/dm_{t\bar t} dp^{t\bar t}_T$          & CMS        &~\cite{Sirunyan:2017azo}               & 16   & 2.01 \\
\hline                                                                                                 
 $d^2\sigma/dm_{t\bar t} d\Delta \eta_{t\bar t}$  & CMS        &~\cite{Sirunyan:2017azo}               & 12   & 1.70 \\
\hline                                                                                                 
 $d^2\sigma/dm_{t\bar t} dy_t$                    & CMS        &~\cite{Sirunyan:2017azo}               & 16   & 1.28 \\
\hline                                                                                                 
 $d^2\sigma/dm_{t\bar t} dy_{t\bar t}$            & CMS        &~\cite{Sirunyan:2017azo}               & 16   & 1.27 \\
\hline
\end{tabular}
\end{center}
\caption{Number of data points and $\chi^2/$Npts for incl. jet and top-quark pair data,
after ePump updating from the CT14HERA2mjet PDFs. } 
\label{Tab:chi2}
\end{table}
%
With direct implementation of ePump updating, we see no 
significant impact from the 1D $t\bar{t}$ distributions on modifying the   
CT14HERA2 PDFs except some minor change  on gluon PDF in the large-$x$ 
region. 
This simply means that the gluon PDF,  in the $x$ range relevant to the 1D $t\bar{t}$ distributions, is already constrained 
by some other data in the original CT14HERA2 fit. 
As shown in Ref.\cite{Hou:2019gfw}, in the framework of CT14HERA2, the gluon PDFs are mainly 
constrained by DIS and jet data. 
In order to see the impact on gluon PDF from $t\bar{t}$ production,
we need to suppress the contribution from jet data. 
For this purpose, the Hessian eigenvector sets CT14HERA2mjet 
are generated from a global fit by including all the 
data used in the CT14HERA2 fit except  the four inclusive 
jet production data from the Tevatron and the LHC Run I.

\begin{figure}[htbp]
    \begin{center}
    \includegraphics[width=0.32\textwidth]{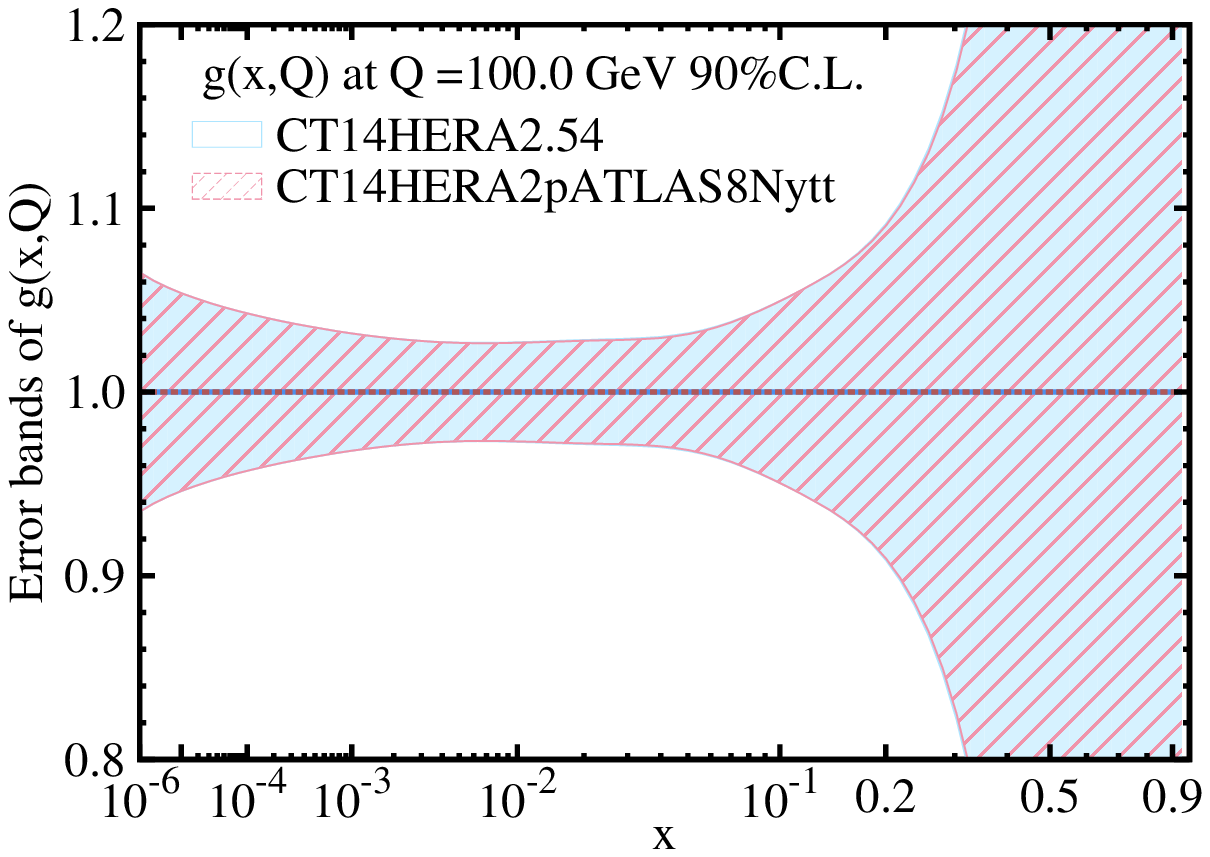} 
    \includegraphics[width=0.32\textwidth]{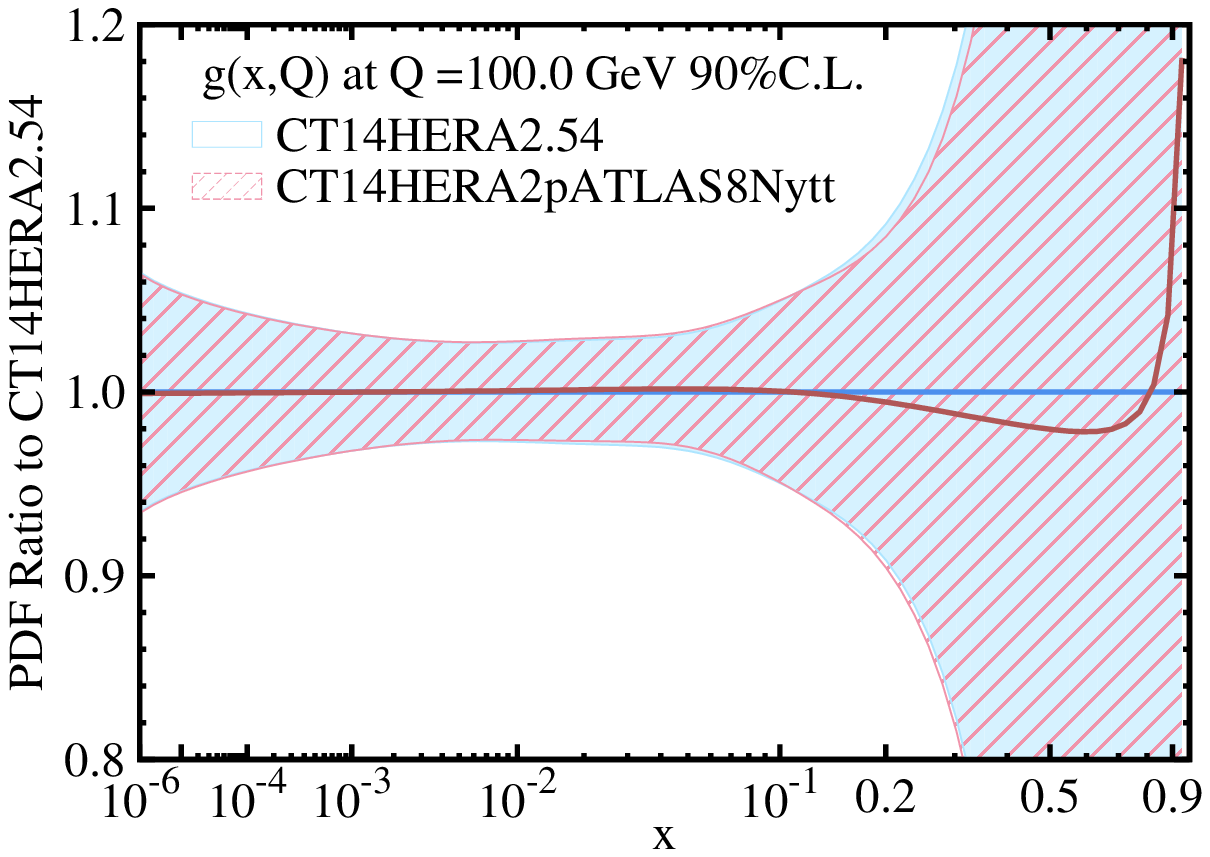} \\
    \includegraphics[width=0.32\textwidth]{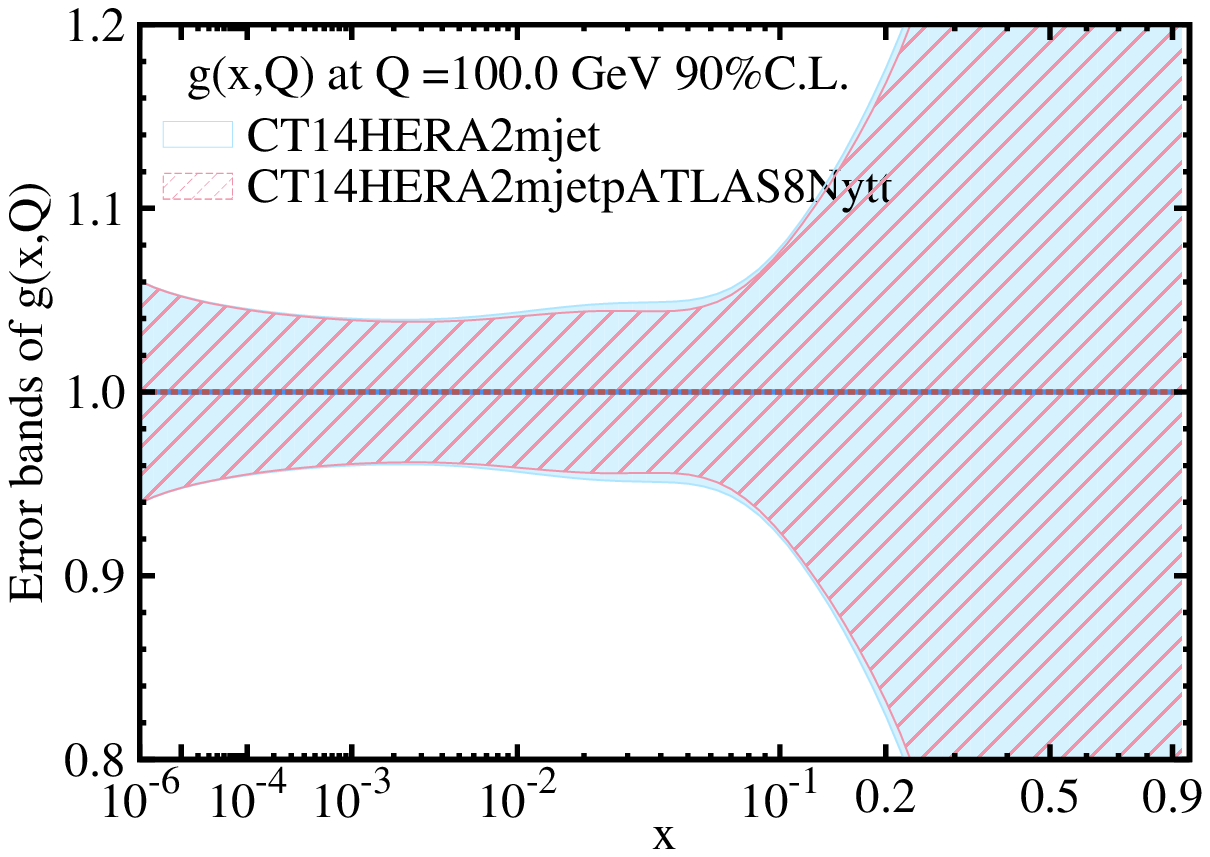}
    \includegraphics[width=0.32\textwidth]{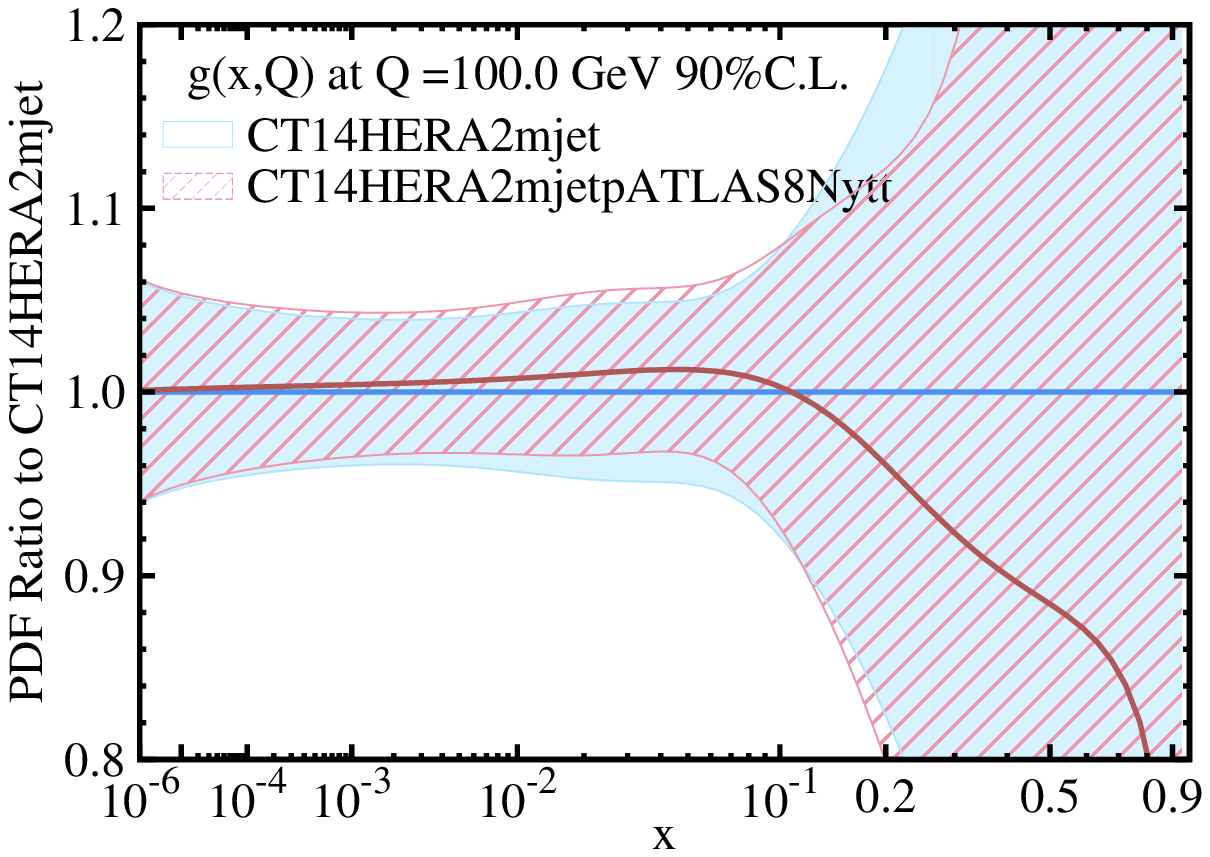} \\
    \includegraphics[width=0.32\textwidth]{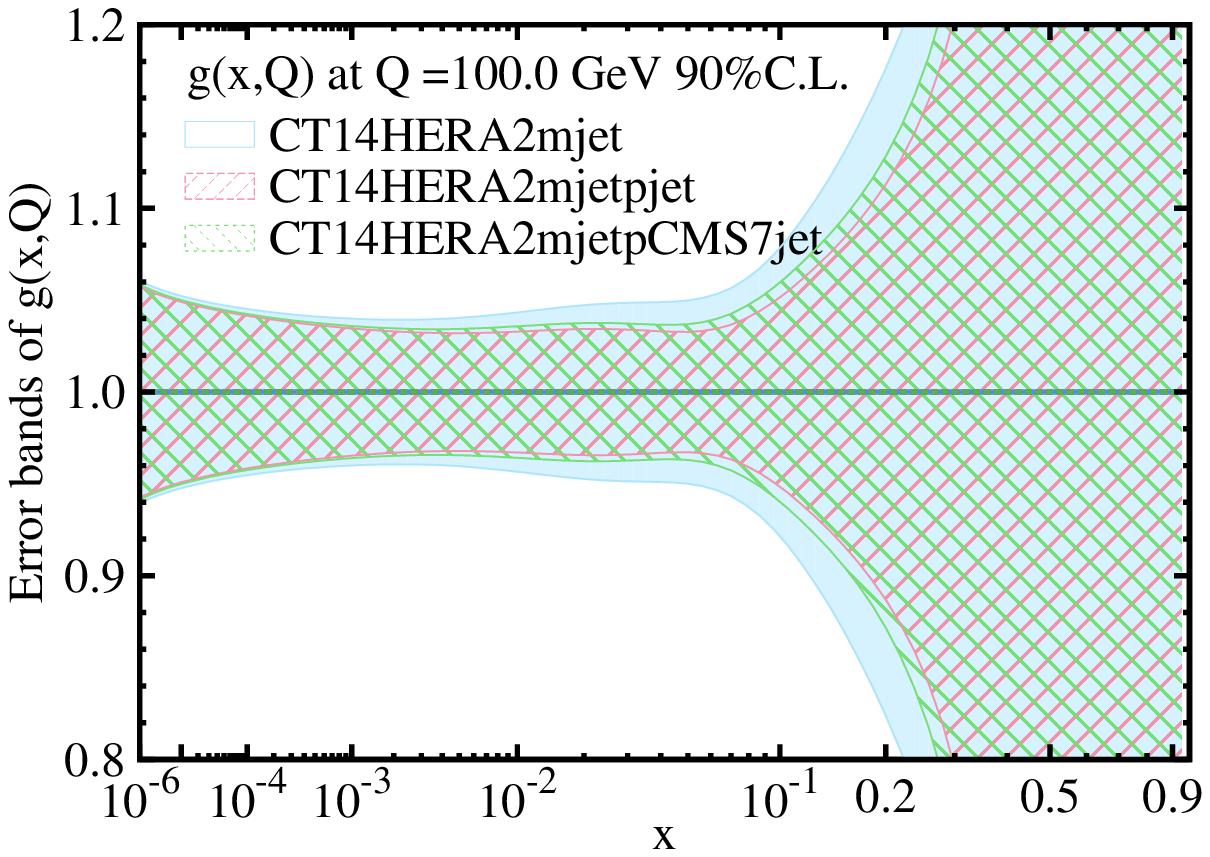}
    \includegraphics[width=0.32\textwidth]{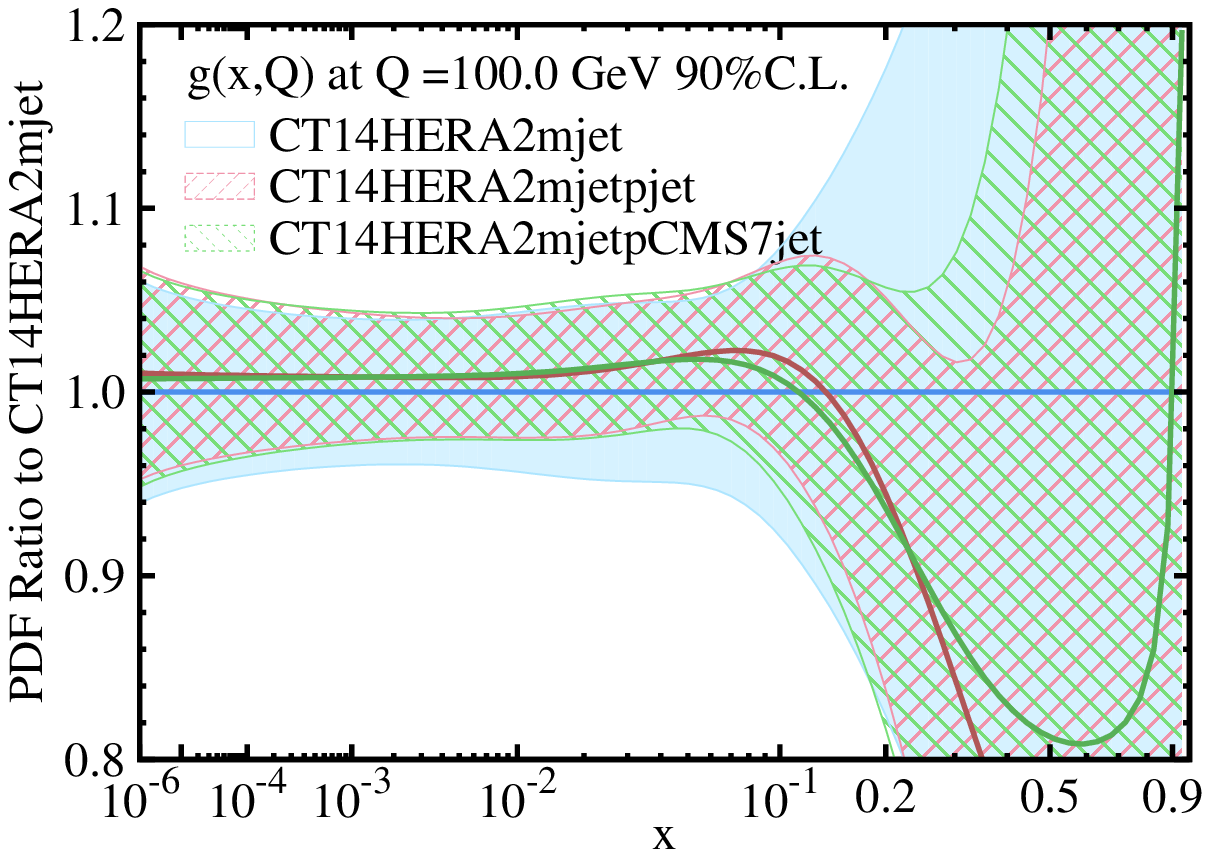} \\ 
    \end{center}
    \caption{ePump updated PDFs, CT14HERA2pATLAS8Nytt and  CT14HERA2mjetpATLAS8Nytt,     
    which are obtained by including normalized ATLAS 8 TeV $y_{t\bar{t}}$ data, 
    are compared with the PDFs before the updating, which are CT14HERA2 and CT14HERA2mjet PDFs, respectively. The CT14HERA2mjetpCMS7jet PDFs are ePump-updated from CT14HERA2mjet by adding only the CMS 7 TeV jet data. 
 }
    \label{Fig:CT14HERA2mjet}
  \end{figure}

Without the jet data included in the starting CT14HERA2mjet PDFs, 
the ePump updated PDFs that include  only the $t\bar{t}$ data in the analysis, receive 
no contribution from jet data.
In Fig.~\ref{Fig:CT14HERA2mjet}, we show both ePump updated PDFs, 
starting from CT14HERA2 and CT14HERA2mjet PDFs by including the
normalized ATLAS 8 TeV $y_{t\bar{t}}$ data.
The impact on gluon PDF from $t\bar{t}$ data can be seen by 
comparing the difference between the gluon PDF before and 
after the ePump updating.
We note that, we do not include the PDF errors induced 
by the two extreme $g$-PDF sets of CT14HERA2 PDFs in this work, for fair comparison of various PDF error sets. 
It is obvious that, without the jet data included in the global analysis, 
the normalized $t\bar{t}$ data have rather 
obvious impact on both the central predictions and uncertainty 
bands of the CT14HERA2mjet PDFs. Hence, the $t\bar{t}$ data can indeed constrain the $g$-PDF in the large-$x$ region. 

The ePump-updated CT14HERA2mjet gluon-PDFs after adding all those four jet data (named 
CT14HERA2mjetpjet) and adding only CMS 7 TeV jet data (named CT14HERA2mjetpCMS7jet) 
are also compared in Fig.~\ref{Fig:CT14HERA2mjet}. 
We first observe that, the CT14HERA2mjetpjet
gluon PDF has much smaller uncertainty band than the 
CT14HERA2mjetpATLAS8Nytt gluon PDF for 
$x$ between 0.01 and 0.3, which shows the much 
stronger constrain on the gluon PDF uncertainty from the 
jet data.
It is therefore understandable why we did not see significant 
impact on the ePump-updated CT14HERA2 PDF by including the $t\bar{t}$ data.
Despite the noticeable difference between the uncertainty bands of  
CT14HERA2mjetpjet and CT14HERA2mjetpATLAS8Nytt gluon PDFs, 
it is worth noting that both the $t\bar{t}$ and jet data 
constrain the central-fit $g$-PDF in a similar way. They all prefer softer gluon in the large-$x$ region, as compared to the CT14HERA2mjet fit. 

%
%
Fig.~\ref{Fig:CT14HERA2mjet} also shows that the CMS 7 TeV inclusive jet data provide the strongest constraint  on the $g$-PDF 
among the four jet data included in the CT14HERA2 fit. 

Below, we explain why the $g$-PDF error band of CT14HERA2mjetpATLAS8Nytt 
is not as narrow as that of CT14HERA2mjetpCMS7jet in the large-$x$ region. Namely, we would explain why jet data provide stronger constraint on $g$-PDF uncertainties than the considered $t\bar{t}$ data.

First, we note that the $t\bar{t}$ data 
have rather smaller number of data points than the jet data,  
by about a factor of 10. 
In Table 1, 
we show the number of data points ($N$)  
for jet data that are included in the CT14HERA2 fit and 
for the new LHC $t\bar{t}$ data.
The values of $\chi^2/N$ in the 
Table 1
are calculated by using ePump to update the CT14HERA2mjet 
PDFs with the inclusion of each individual data set. 
As discussed above, the sensitivity of the $t\bar{t}$ data to $g$-PDF is not as large as the jet data, to  constrain the $g$-PDF uncertainties in the large-$x$ region. 
Nevertheless, 
 it is also interesting to compare the sensitivity per data point of  
the jet and $t\bar{t}$ data. 
In order to see this, a hypothetical weight is assigned  
to the 1D $t\bar{t}$ distribution data with the weight 
equal to the ratio between the number of data points of the
CMS 7 TeV jet data and the considered $t\bar{t}$ distribution. 
Taking the normalized CMS 8TeV $p_T$ distribution as an example, 
the hypothetical weight that applies to the data is equal to 
$w = 133/8 = 16.6$. 
In practice, a larger weight can arise from increasing the 
event statistics (e.g., with a larger integrated collider luminosity) or reducing the experimental errors (e.g., with improvement in detection efficiency). 
%
  \begin{figure}[h]
    \begin{center}
    \includegraphics[width=0.32\textwidth]{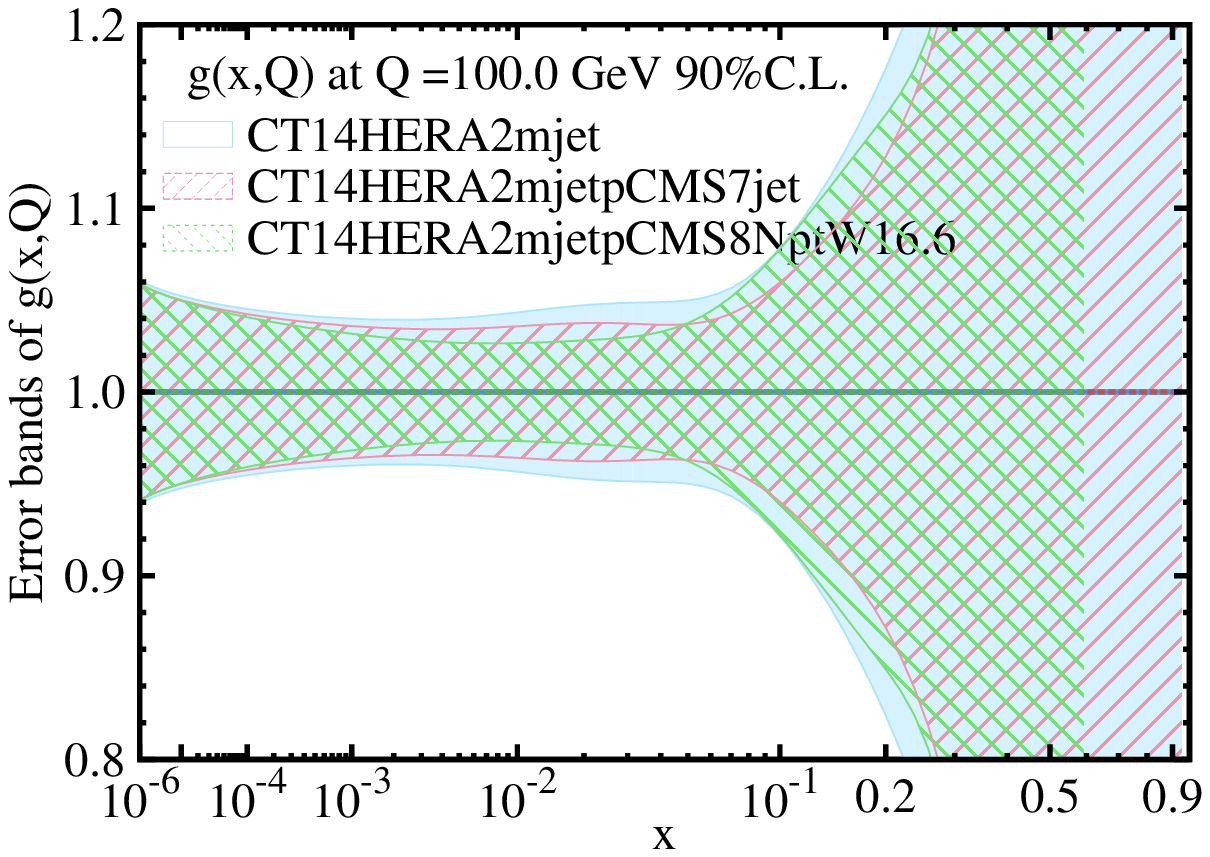} 
    \includegraphics[width=0.32\textwidth]{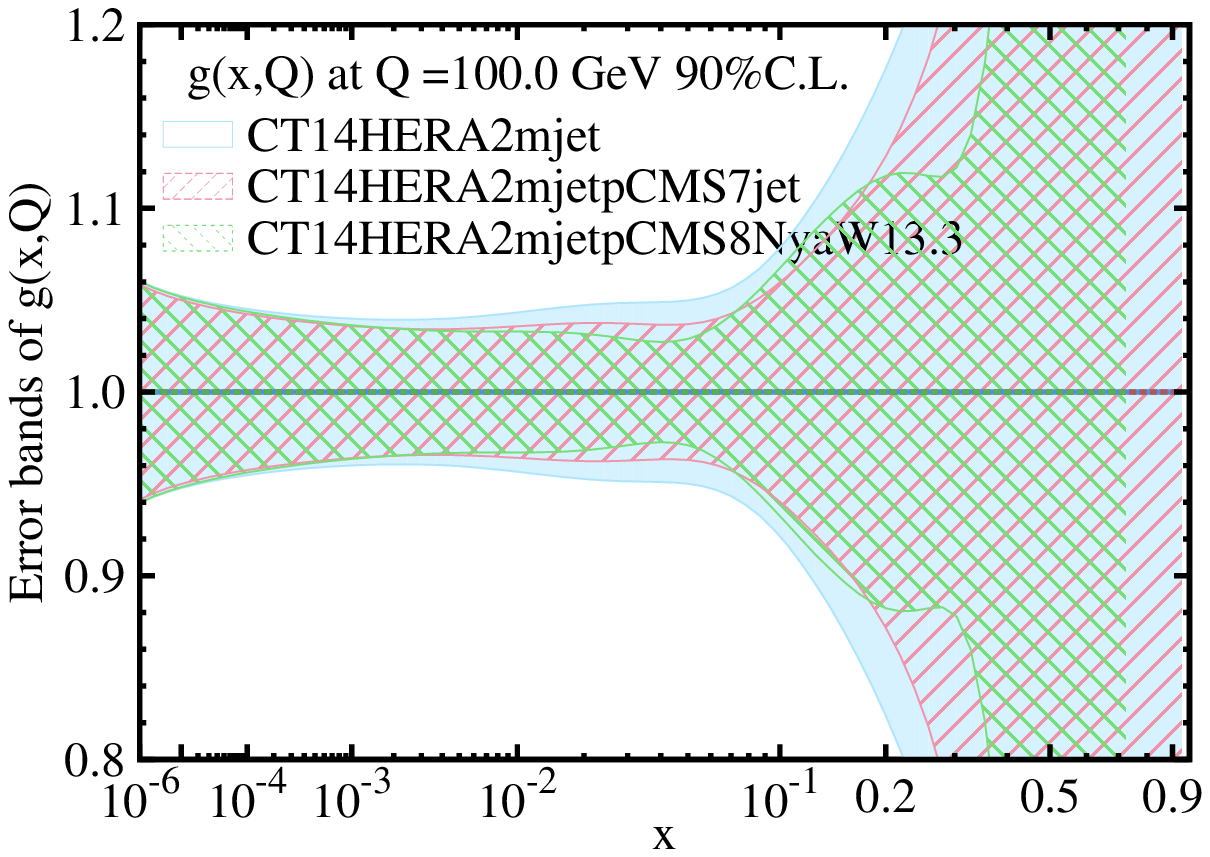}  \\ 
    \includegraphics[width=0.32\textwidth]{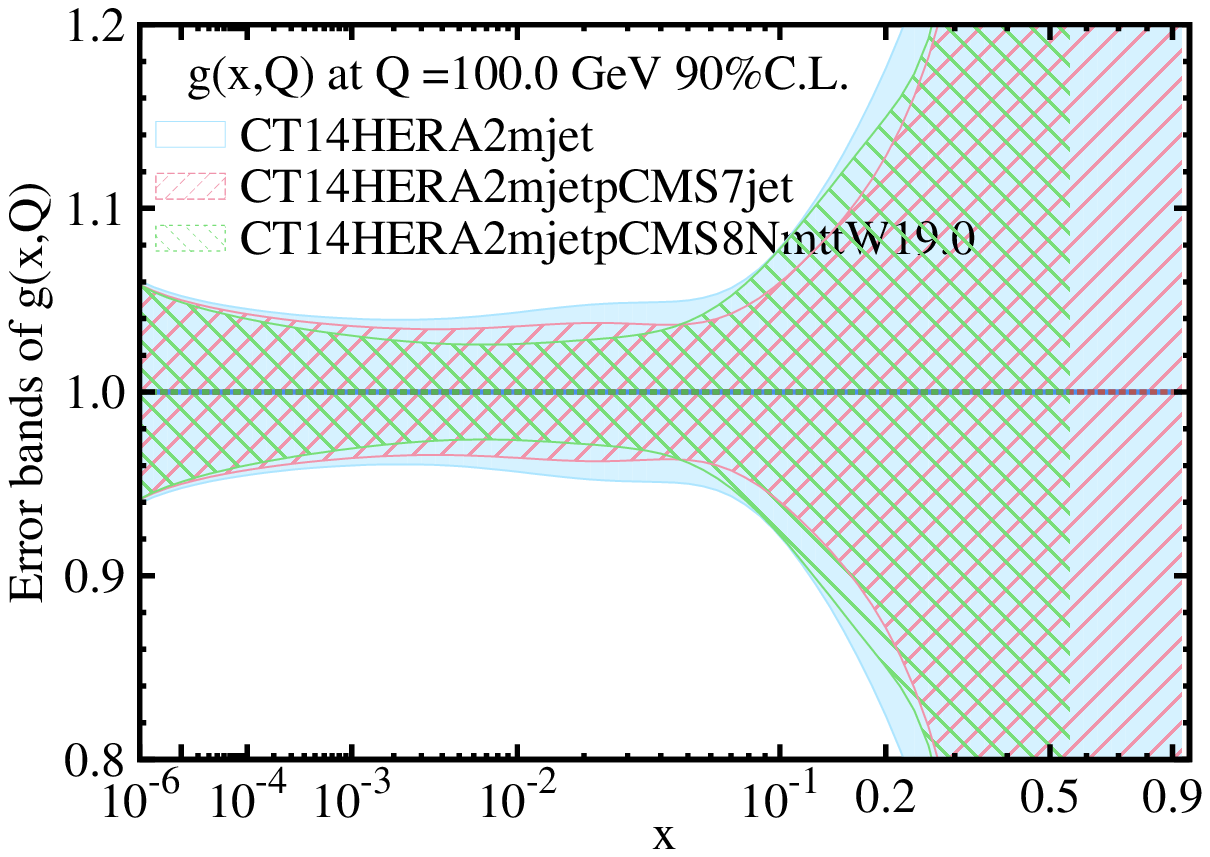} 
    \includegraphics[width=0.32\textwidth]{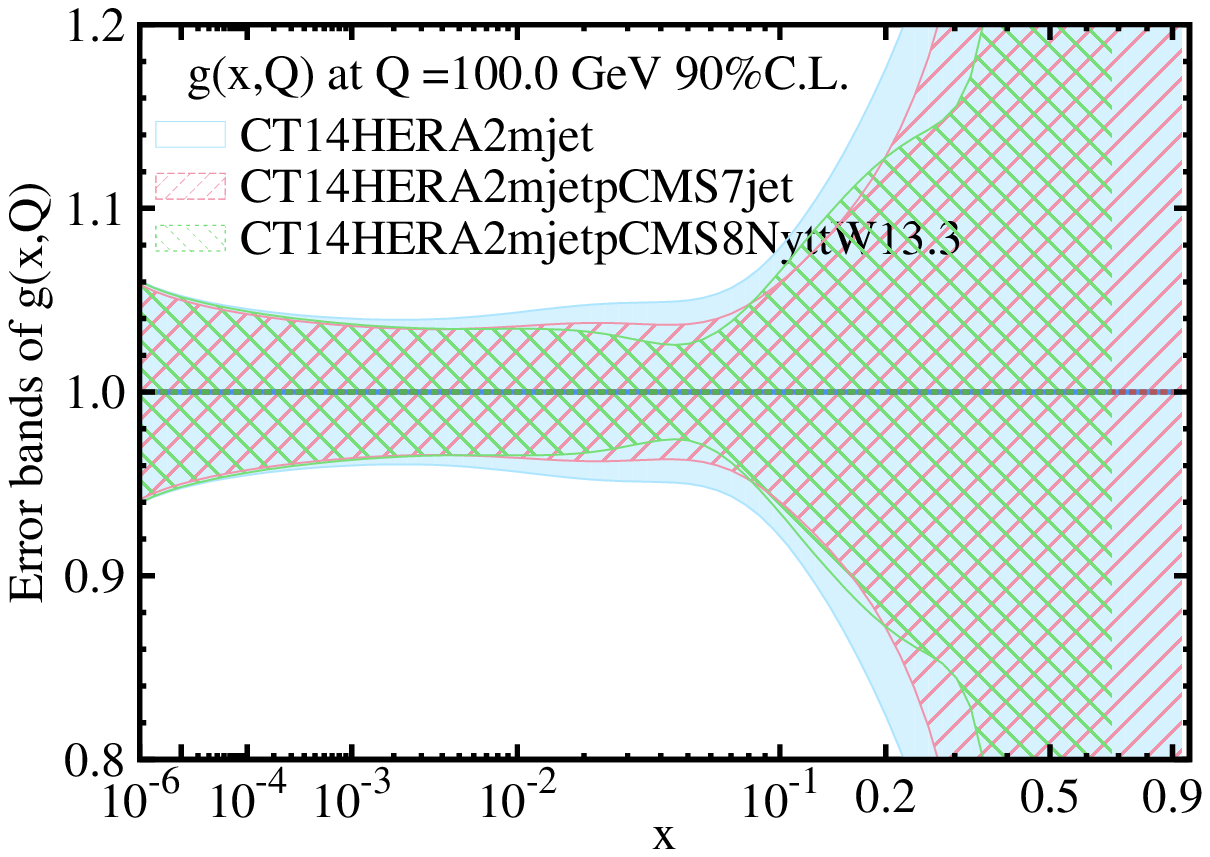} \\
    \end{center}
    \caption{Comparison of CT14HERA2mjet and ePump-updated PDFs, at Q = 100 GeV and at 90\% C.L., by adding only the  
             CMS 7 TeV jet data or the normalized CMS 8 TeV 1D $t\bar{t}$ 
             data, by adding one at a time, with hypothetical weights for various $t\bar{t}$  distributions. }
    \label{Fig:1DnCMS}
  \end{figure}
%
In this naive estimation, we assume the central values 
of the measurement do not change so that we only show the comparison on the PDF uncertainties in the following figures.

\begin{figure}[h]
    \begin{center}
    \includegraphics[width=0.32\textwidth]{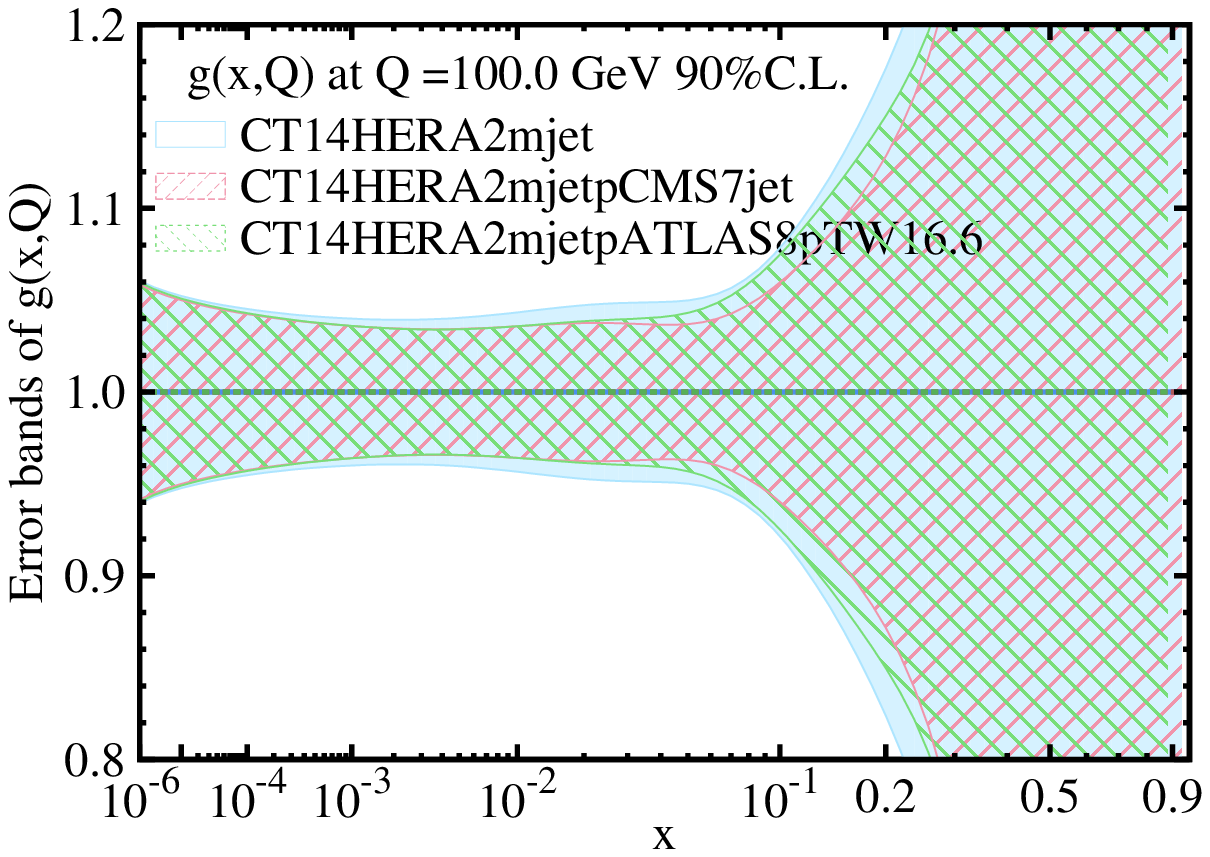} 
    \includegraphics[width=0.32\textwidth]{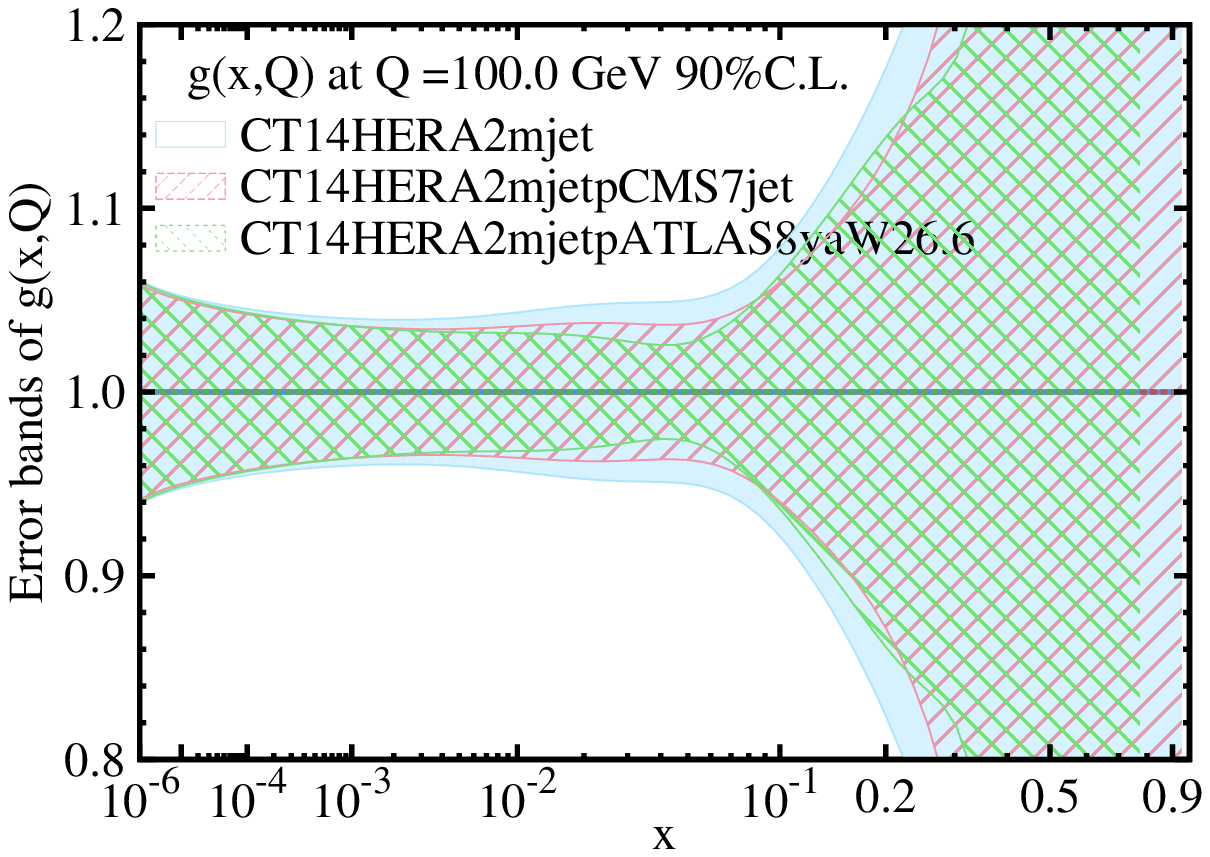}   \\
    \includegraphics[width=0.32\textwidth]{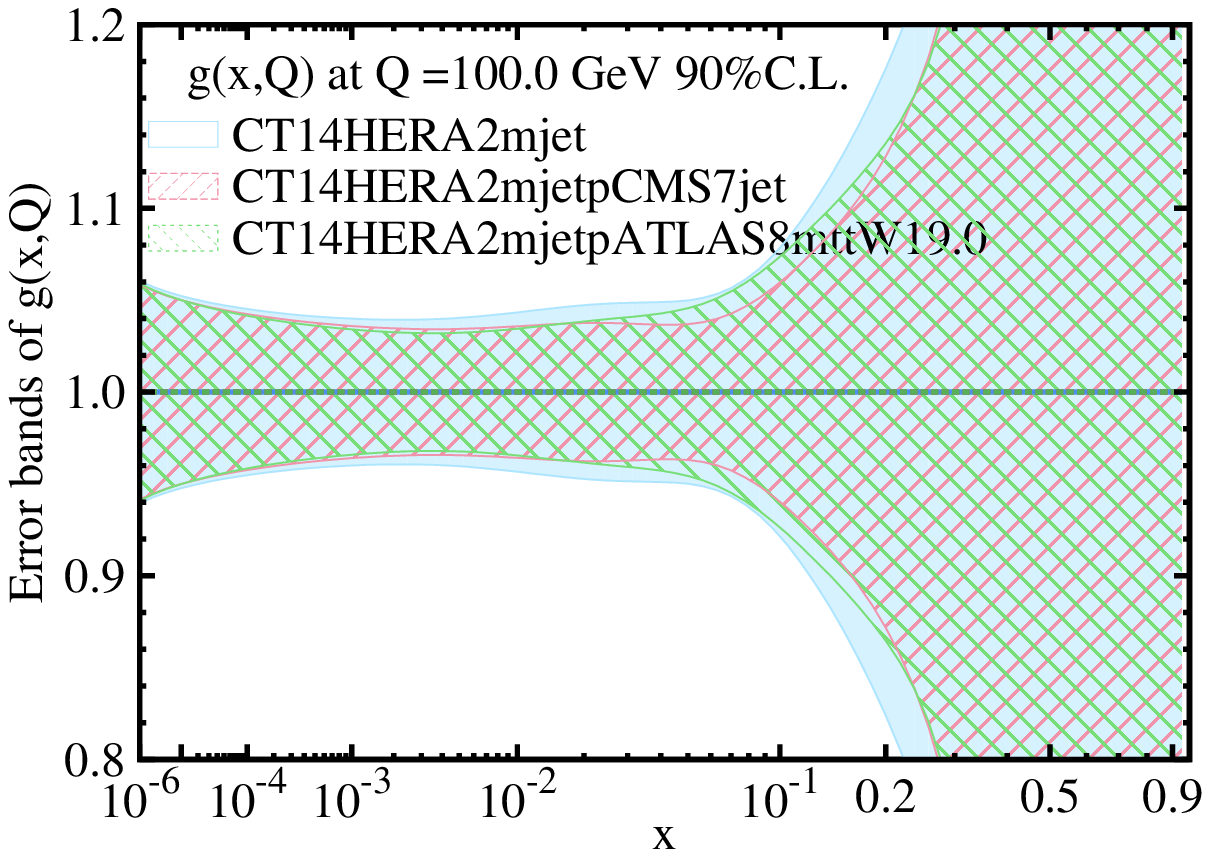} 
    \includegraphics[width=0.32\textwidth]{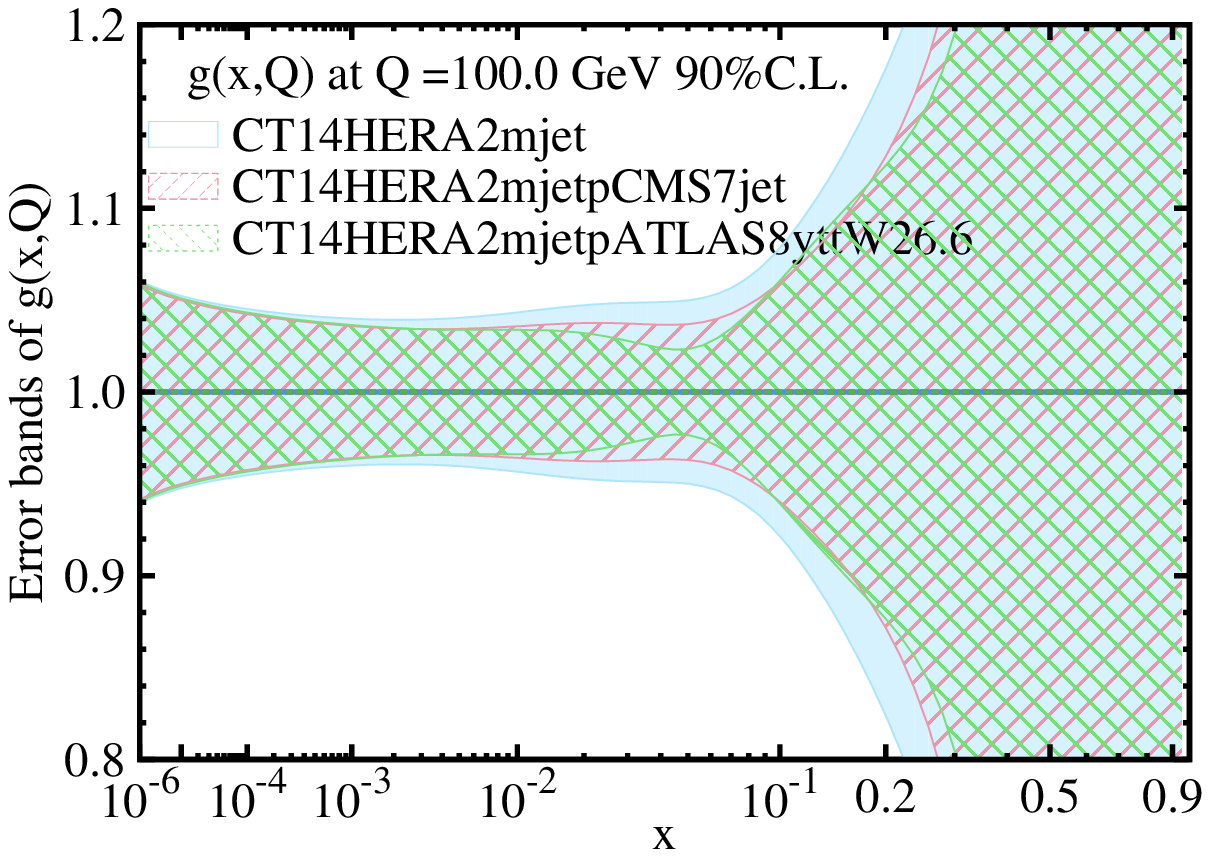} \\
    \end{center}
    \caption{Comparison of CT14HERA2mjet and ePump-updated PDFs,  at Q = 100 GeV and at 90\% C.L., by adding the  
             CMS 7 TeV jet data and the absolute CMS 8 TeV 1D $t\bar{t}$   
             data, by adding one at a time, with hypothetical weight for various $t\bar{t}$ distributions.  }
    \label{Fig:1DaATL}
  \end{figure}

In Fig.~\ref{Fig:1DnCMS}, we compare the impact on 
$g$-PDF uncertainty from the CMS 7 TeV jet 
data and the normalized CMS 8TeV 1D $t\bar{t}$ 
distribution data, with the hypothetical weight discussed above.
It shows that, the weighted $t\bar{t}$ distribution data provide stronger 
constraint on gluon PDFs for $10^{-3} \lesssim x \lesssim 5 \times 10^{-2}$. 
This conclusion also holds for the absolute ATLAS 8 TeV 1D 
$t\bar{t}$ distribution data. 
With the hypothetical weight equal to the ratio of the number of jet 
and $t\bar{t}$ data points, 
the absolute 1D $t\bar{t}$ distribution data provide about the same 
constraint on gluon PDF as the jet data, which is shown in 
Fig.~\ref{Fig:1DaATL}. 

\begin{figure}[htbp]
    \begin{center}
    \includegraphics[width=0.32\textwidth]{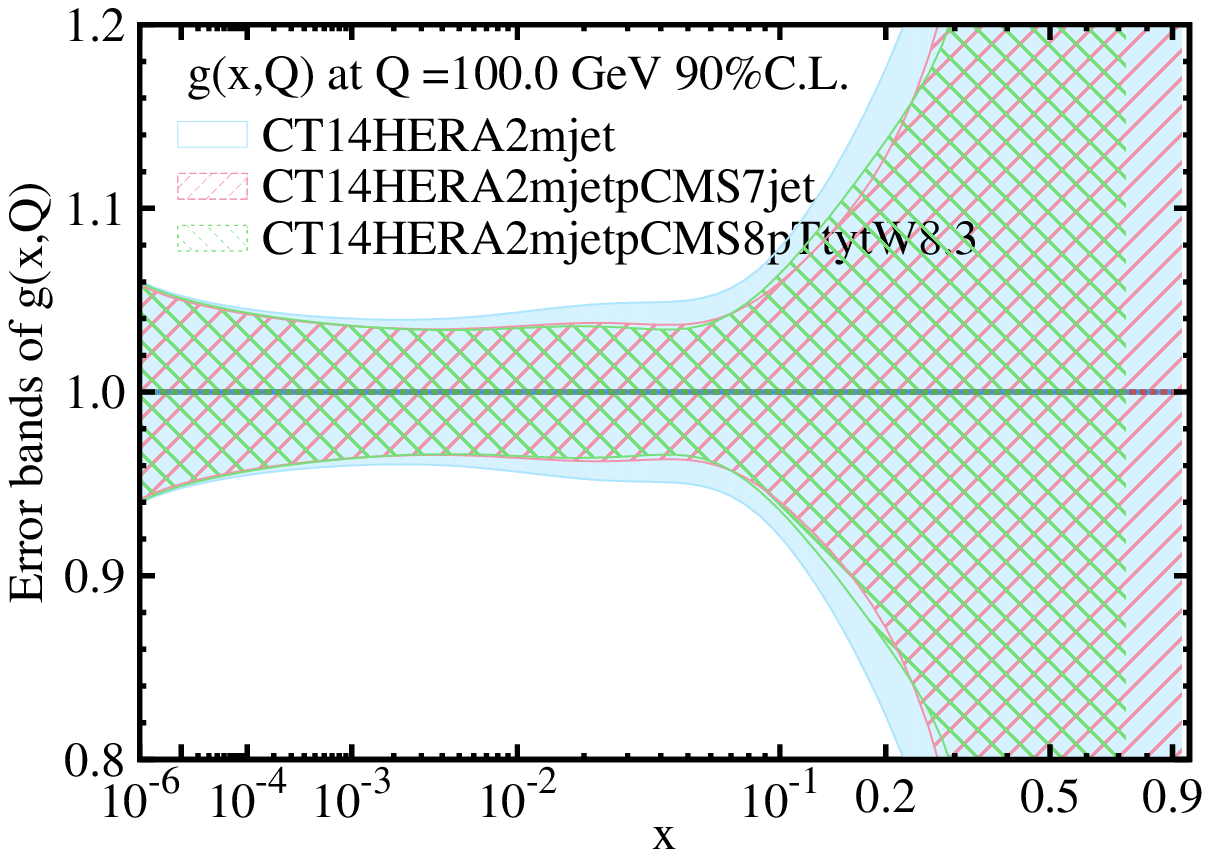} 
    \includegraphics[width=0.32\textwidth]{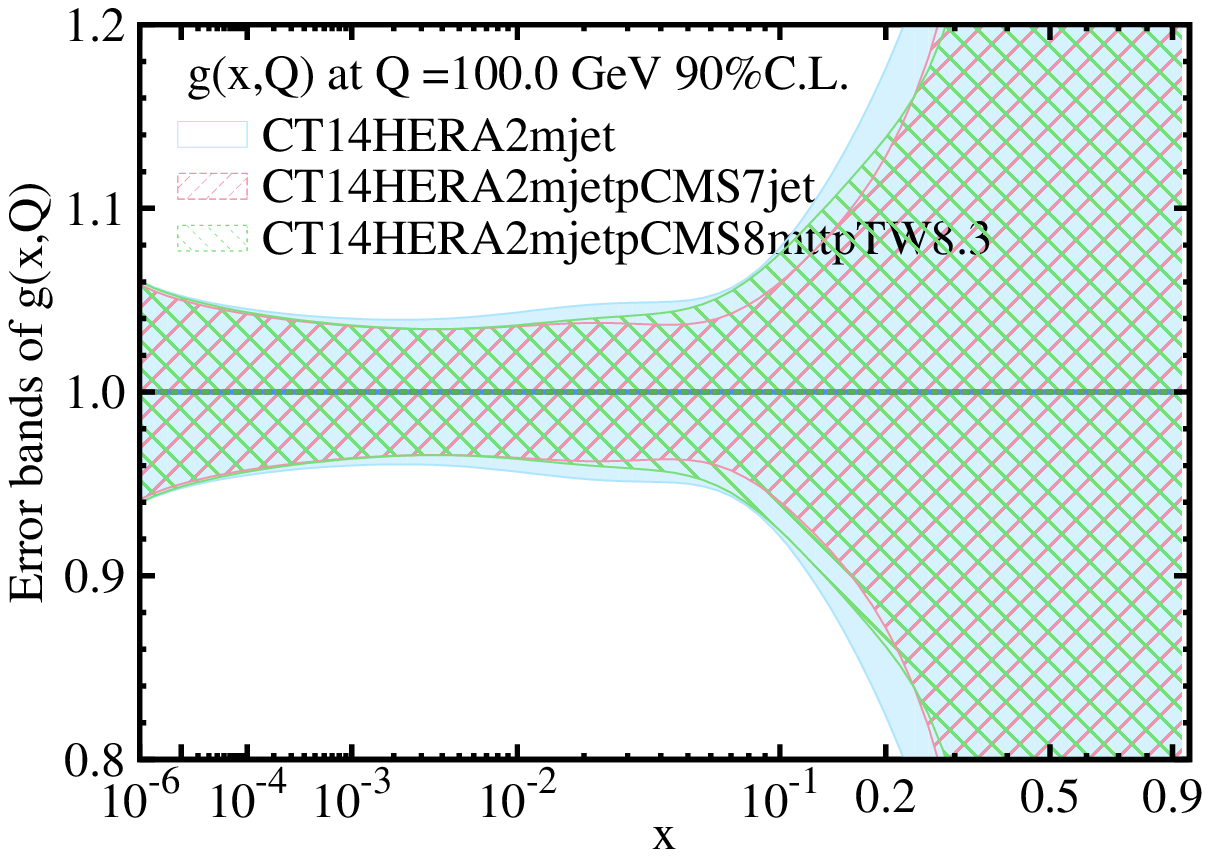} 
    \includegraphics[width=0.32\textwidth]{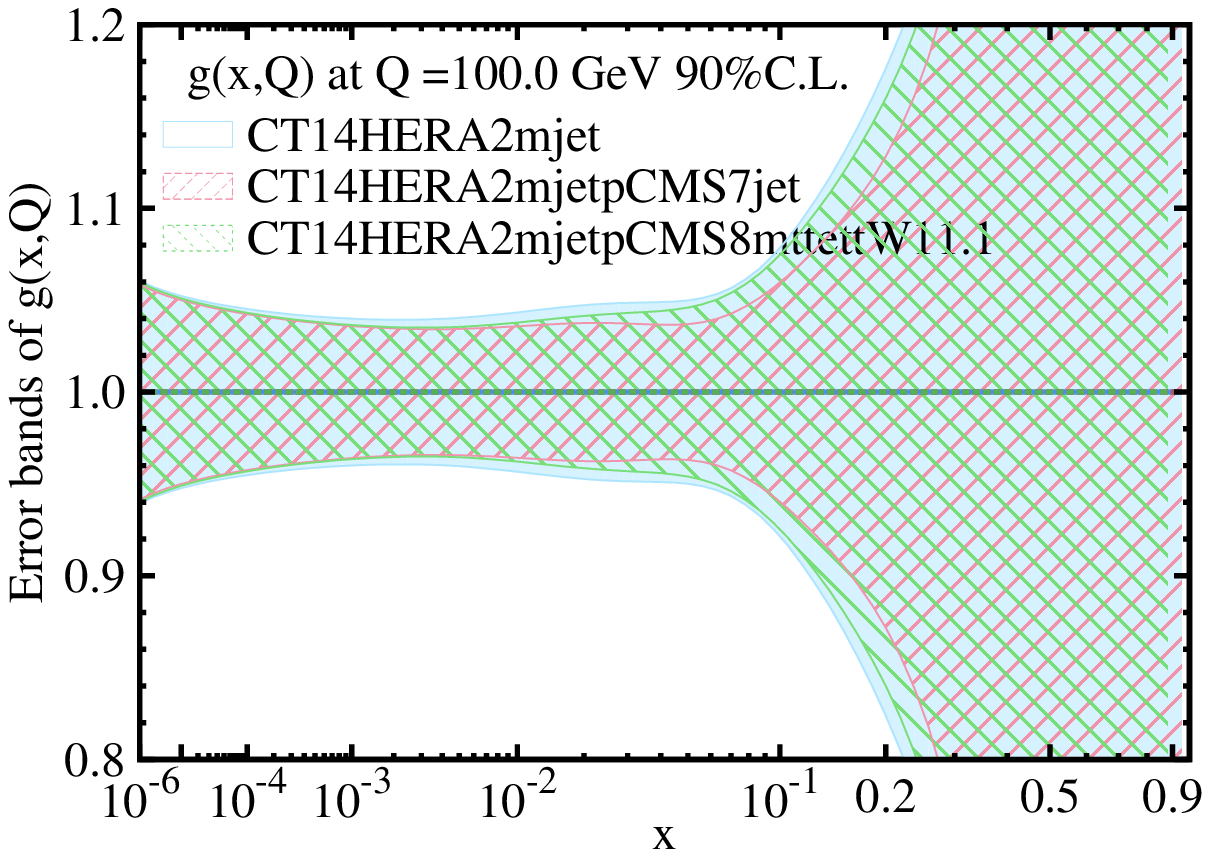} 
    \includegraphics[width=0.32\textwidth]{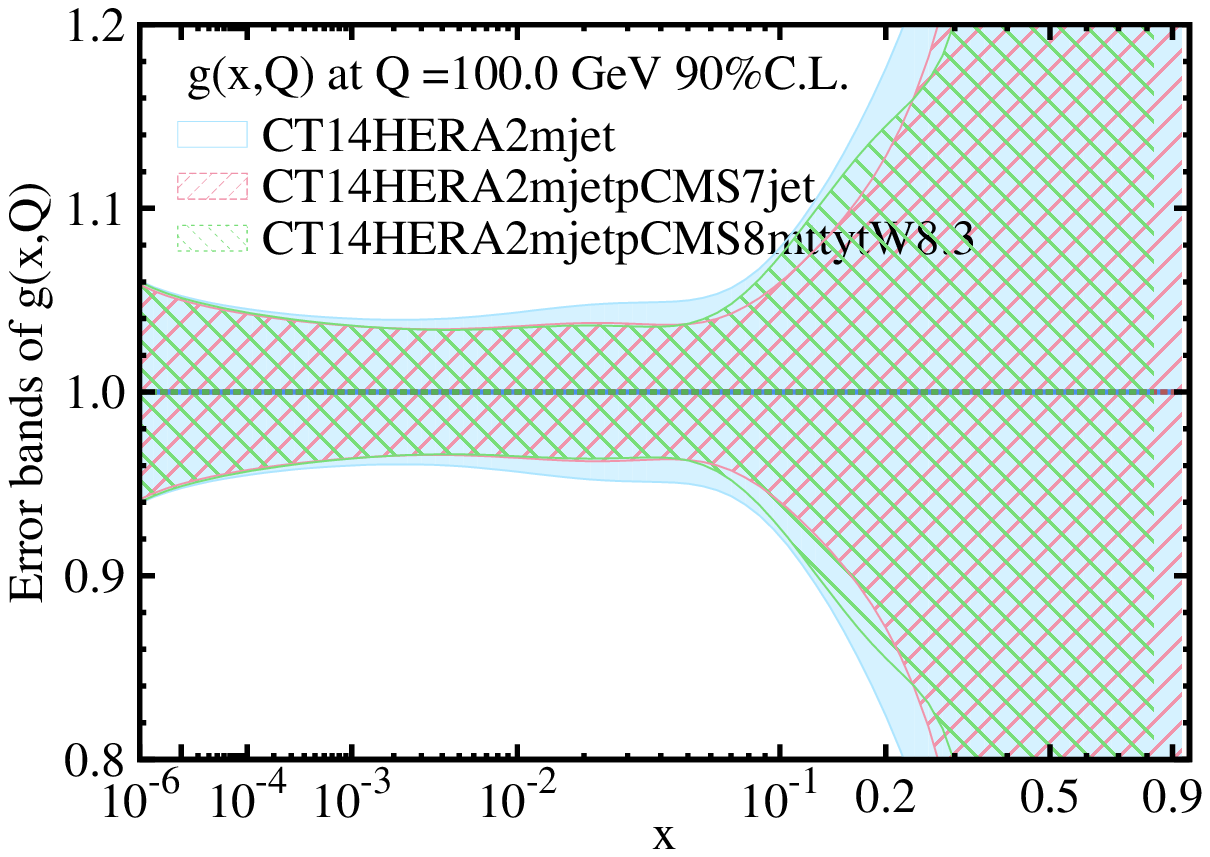} 
    \includegraphics[width=0.32\textwidth]{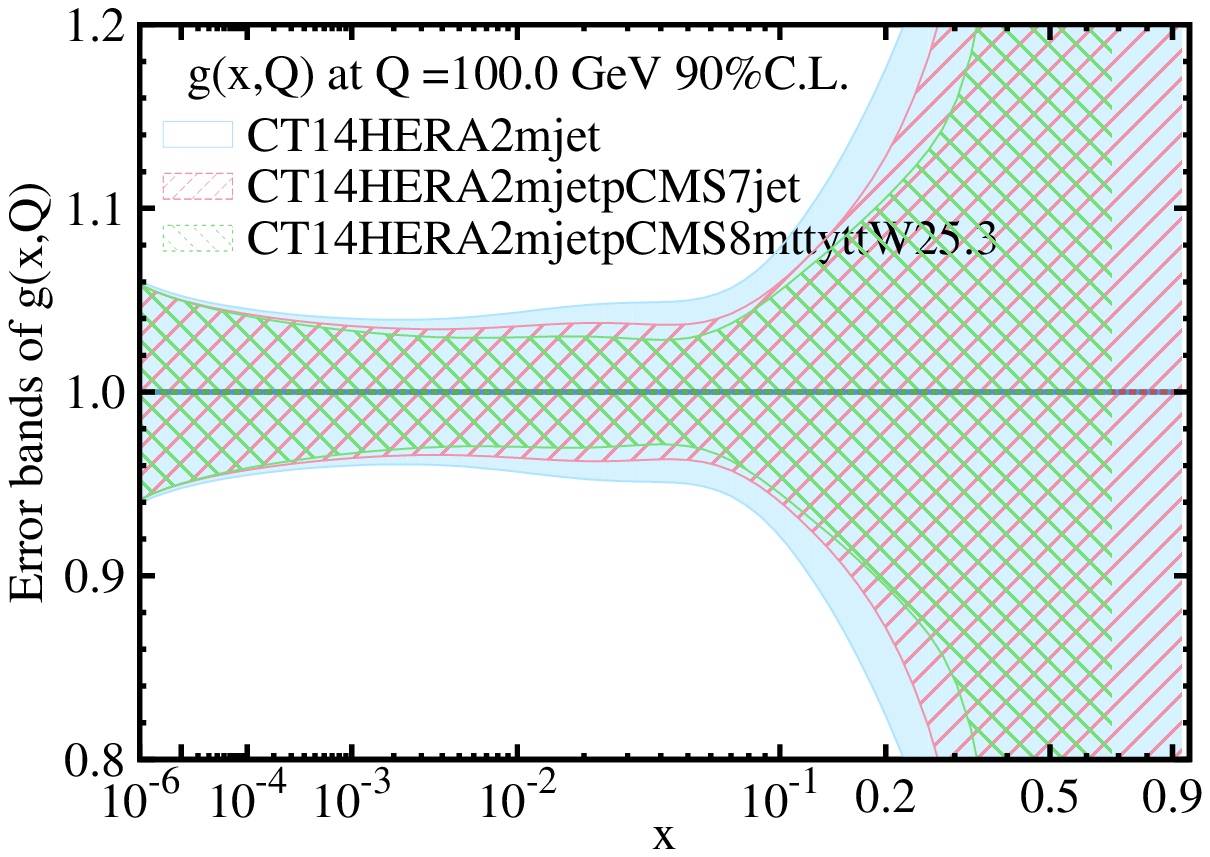} 
    \end{center}
    \caption{ Comparison of the impact on gluon uncertainty between CMS 7 TeV jet data
              and the absolute CMS 8TeV 2D $t\bar{t}$ data with a hypothetical weight 
              which equals to the ratio of the number of jet and $t\bar{t}$ data points, for the considered distribution. }
    \label{Fig:2Dttb}
  \end{figure}

Further examination on the absolute CMS 8 TeV two-dimentional 
$t\bar{t}$ distribution data also shows no significant impact on ePump-updating  
CT14HERA2 and CT14HERA2mjet PDFs. Similar to the 1D $t\bar{t}$ 
data, the 2D $t\bar{t}$ data also show compatible sensitivity to updating the CT14HERA2mjet gluon PDF as the jet data, when a hypothetical weight 
is assigned to equal to the ratio of the number of jet and $t\bar{t}$ data points, for the considered distribution.
The result of comparison is shown in Fig.~\ref{Fig:2Dttb}.

\begin{figure}[htbp]
    \begin{center}
    \includegraphics[width=0.32\textwidth]{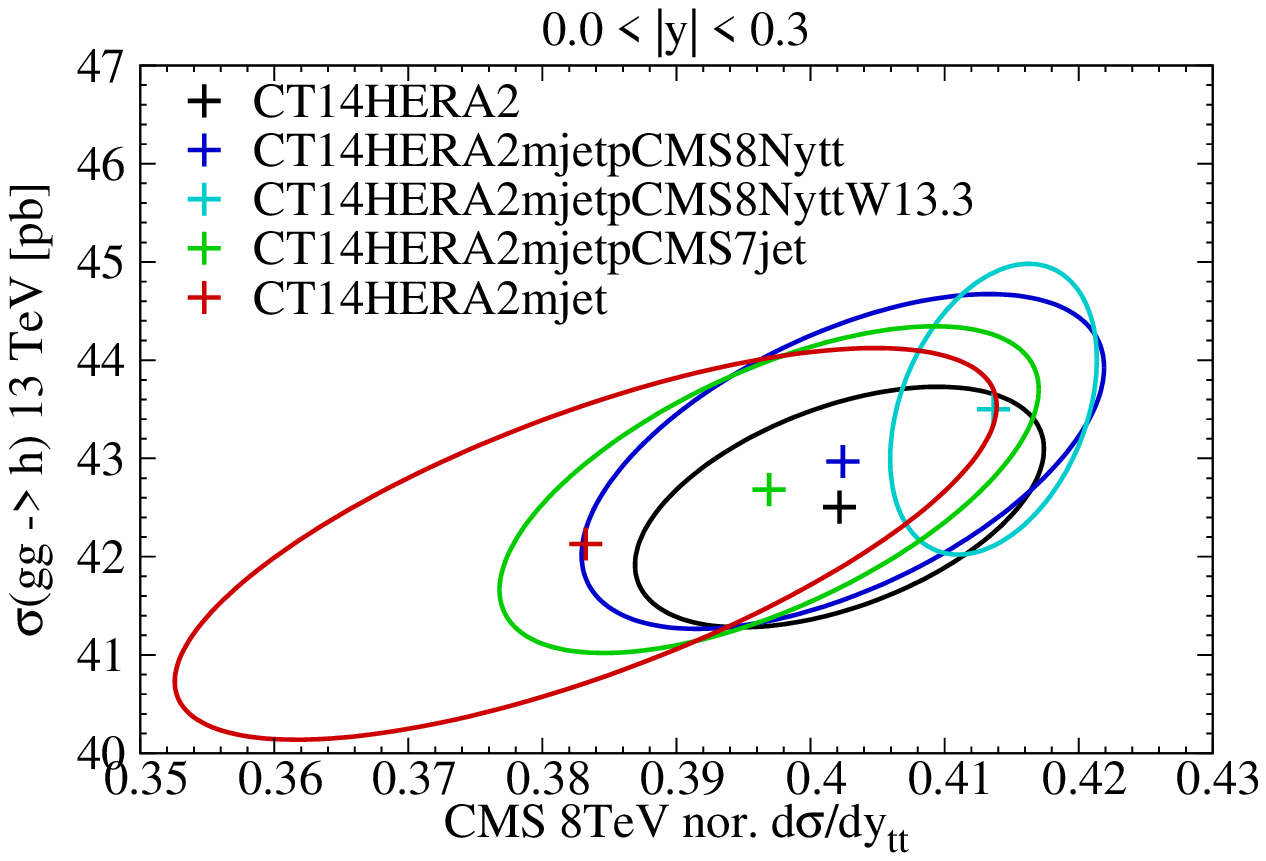} 
    \includegraphics[width=0.32\textwidth]{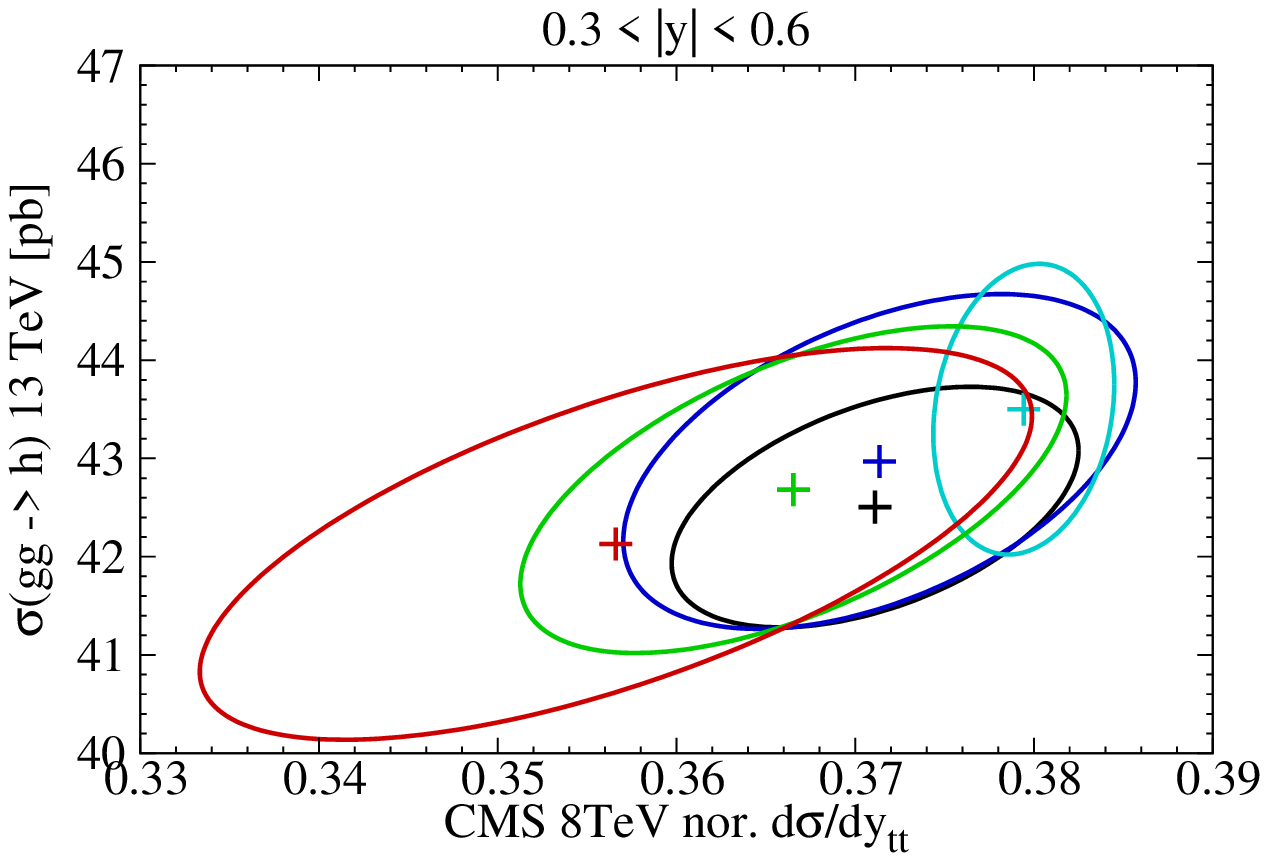}
    \includegraphics[width=0.32\textwidth]{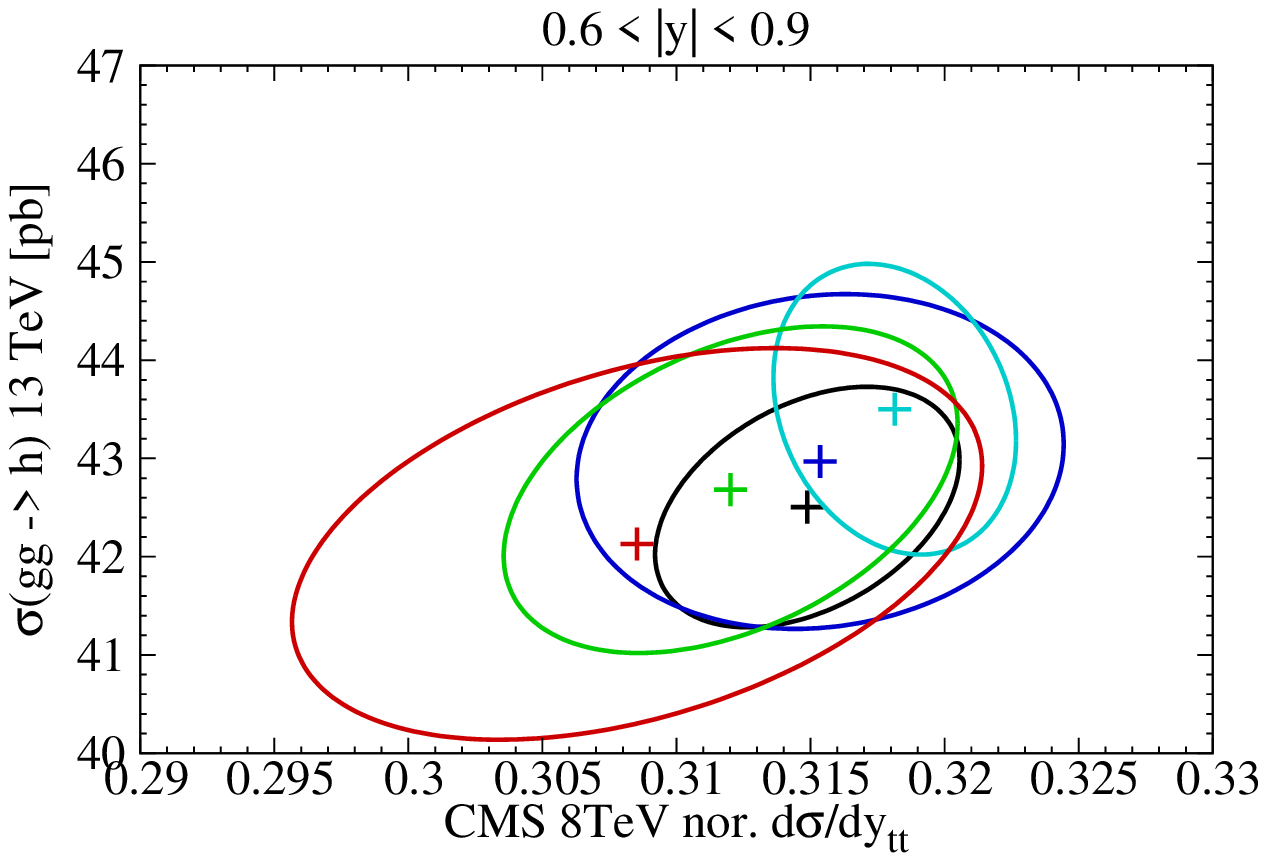} \\
    \includegraphics[width=0.32\textwidth]{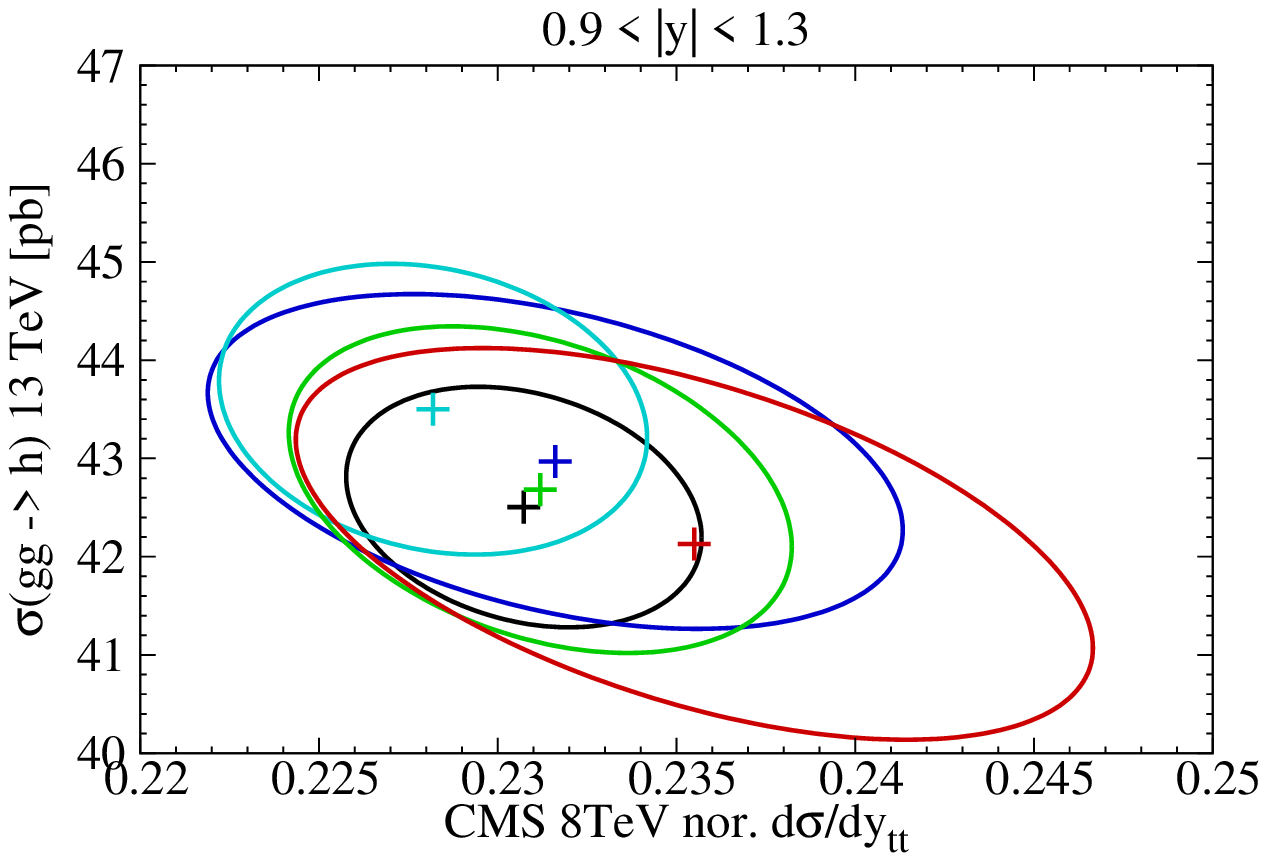}
    \includegraphics[width=0.32\textwidth]{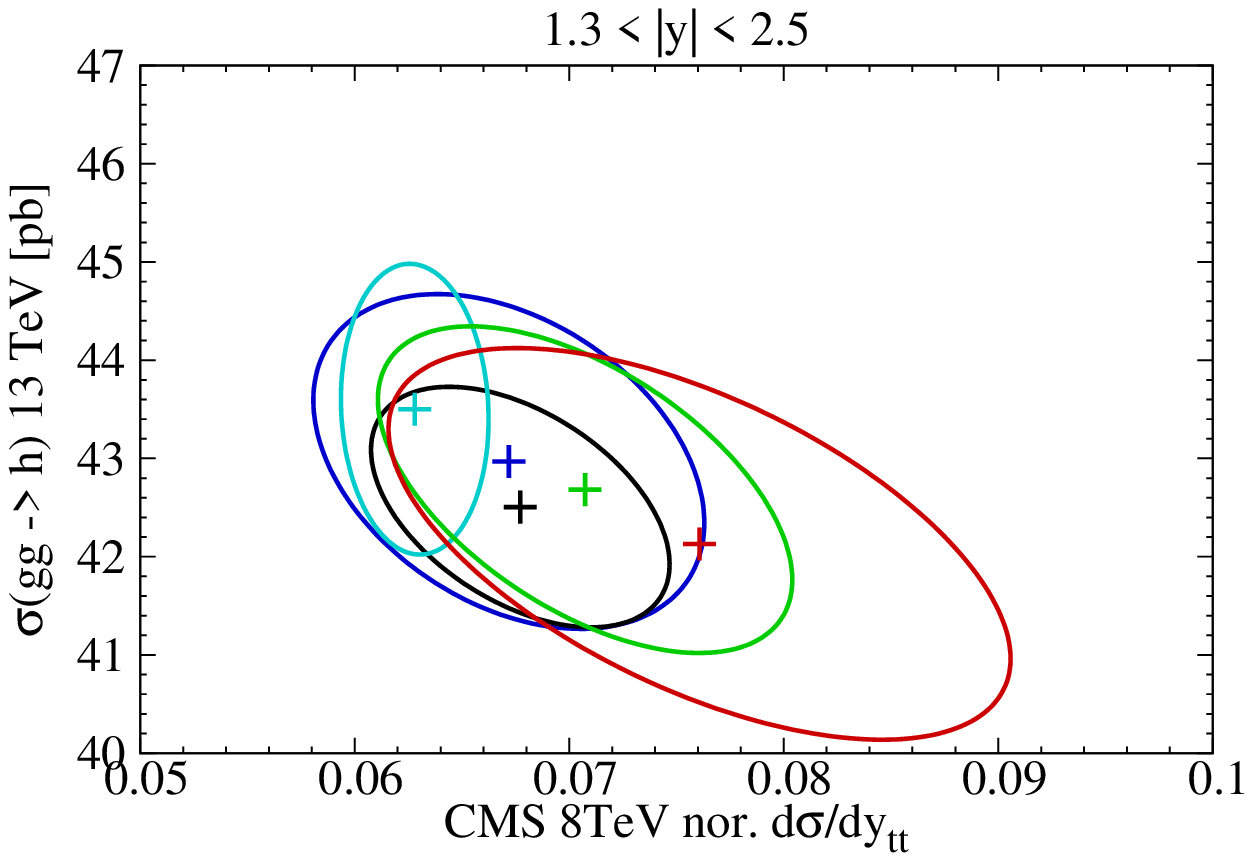} \\
    \end{center}
    \caption{ Correlation ellipse between CMS 8 TeV normalized $y_{t\bar{t}}$ 
    data for various rapidity bins and Higgs production through gluon-gluon fusion 
    at the 13 TeV LHC for CT14HERA2 (black), CT14HERA2mjetpCMS8Nytt (blue), 
    CT14HERA2mjetpCMS8NyttW13.3 (cyan), CT14HERA2mjetpCMS7jet (green) and 
    CT14HERA2mjet (red). 
    The central prediction of the CT14HERA2mjetpCMS8NyttW13.3 is obtained by assuming 
    the central measurement is the same as that in CT14HERA2mjetpCMS8Nytt.
    }
    \label{Fig:ggh}
  \end{figure}

Next, we examine the impact of the updated PDFs, 
obtained by including various $t\bar{t}$ data in the ePump updating.  
The Higgs production rate through gluon-gluon fusion at 
the LHC is sensitive to $g$-PDF in the middle-$x$ region, 
which is constrained by both the jet and $t\bar{t}$ data. 
In Fig.~\ref{Fig:ggh}, we show the correlation ellipses of the Higgs 
production rate via gluon-gluon fusion and the CMS 8 TeV 
normalized $y_{t\bar{t}}$ differential cross section 
(with weight 1 or 13.3, respectively), 
for various ePump updating scenarios.  

%
%
In summary, we observe that the present top-quark pair production data 
show minor impact on updating the CT14HERA2 gluon PDF where some Tevatron and LHC jet data have already been included 
in the global analysis. 
This is because the number of data points for the $t\bar{t}$ 
data is much less than the jet data.
Though the overall sensitivity of the present $t\bar{t}$ data is smaller than the jet data, the $t\bar{t}$ data constrain the central-fit $g$-PDF in the same way as the CMS 7 TeV jet data. Hence, with increasing number of $t\bar{t}$ data collected at the future LHC runs, the $t\bar{t}$ data can provide as strong constrain on $g$-PDF uncertainty as jet data in their common $x$ values. It may even provide stronger constraint than jet data in somewhat larger $x$ values where the theoretical uncertainty of the NNLO $t\bar{t}$ calculation can be smaller than that of the NNLO inclusive jet cross section calculation.
We also showed that the sensitivity per data point of the jet and  $t\bar{t}$ data, for constraining the $g$-PDF in the similar $x$ range, are about the same.
This is done by assigning a hypothetical weight to the $t\bar{t}$ data, as the ratio 
of the number of total data points between jet data and the $t\bar{t}$ data under consideration.  
We find that the weighted $t\bar{t}$ data can constrain $g$-PDF uncertainty as well as the jet data.
Hence, we conclude that the sensitivity per data point of the LHC $t\bar{t}$ data is 
similar to that of jet data, as   included in the CT14HERA2 fit, while the total sensitivity of the  
present $t\bar{t}$ data is not as large as the jet data. This is because the sensitivity of 
the whole data set depends on the total number of data points, and 
the total number of data points of the 
presently available $t\bar{t}$ data is smaller than that of the LHC jet data.

\begin{acknowledgments}
We thank our CTEQ-TEA collaborators for fruitful discussions. 
A.M. thanks the Department of Physics at Princeton University for hospitality during the completion of this work.
The work of M.C. was supported by the Deutsche Forschungsgemeinschaft under grant 396021762 - TRR 257. The research of A.M. and A.S.P. has received funding from the European Research Council (ERC) under the European Union’s Horizon 2020 research and innovation programme (grant agreement No 683211); it is also supported by the UK STFC grants ST/L002760/1 and ST/K004883/1.
A.S.P. is a cross-disciplinary post-doctoral fellow supported by funding from
the University of Edinburgh and Medical Research Council (core grant to
the MRC Institute of Genetics and Molecular Medicine).
The work of C.-P. Yuan was supported by the U.S. National Science Foundation under Grant No. PHY-1719914, and he is also grateful for the support from the Wu-Ki Tung endowed chair in particle physics.

\end{acknowledgments}

\end{document}